\documentclass[aps,preprintnumbers,amsmath,amssymb,floatfix,groupedaddress,nofootinbib]{revtex4}

\usepackage{enumerate}
\usepackage{graphicx}
\usepackage{color}
\usepackage[utf8]{inputenc}
\usepackage{enumitem}
\usepackage{appendix}
\usepackage{ulem}

\usepackage{fontenc}  
\usepackage{amsmath}

\newcommand{\beq}{\begin{equation}}
\newcommand{\eeq}{\end{equation}}
\newcommand{\bea}{\begin{eqnarray}}
\newcommand{\eea}{\end{eqnarray}}
\newcommand{\ba}{\begin{array}}
\newcommand{\ea}{\end{array}}

\newcommand{\bef}{\begin{figure}}
\newcommand{\eef}{\end{figure}}

\begin{document}

\title{The two-spin enigma:
from the helium atom to quantum ontology}

\author{Philippe Grangier$^1$, Alexia Auff\`eves$^2$, Nayla Farouki$^3$, Mathias Van Den Bossche$^4$, Olivier Ezratty$^5$\\}
\vspace{3mm}
\affiliation{$^1$ Laboratoire Charles Fabry, IOGS, CNRS, Universit\'e Paris Saclay, F91127 Palaiseau, France.}
\affiliation{$^2$ MajuLab, CNRS–UCA-SU-NUS-NTU International Joint Research Laboratory, Singapore.}
\affiliation{$^3$ Philosopher and historian of science, previously consultant at CEA-Grenoble, France.}
\affiliation{$^4$ Thales Alenia Space, 26 avenue J.-F. Champollion, 31037 Toulouse, France.}
\affiliation{$^5$ EPITA Research Laboratory, 14-16 rue Voltaire, 94270 Le Kremlin-Bic\^etre, France.}

\begin{abstract}
The purpose of this article is to provide  a novel approach and justification of the idea that classical physics and quantum physics can neither function nor even be conceived one without the other - in line with ideas attributed to e.g. Niels Bohr or Lev Landau. Though this point of view may go against current common wisdom,  we will show that it perfectly fits with empirical evidence, and can be maintained without giving up physical realism. In order to place our arguments in a convenient historical perspective, we will proceed as  if we were following the path of a police investigation, about the demise, or vanishing, of some valuable properties of the two electrons in the helium atom. We will start from experimentally based evidence in order to analyse and explain physical facts, moving cautiously from a classical to a quantum description, without mixing them up.  The overall picture will be that the physical properties of microscopic systems are {\bf quantized}, as initially shown by Planck and Einstein, and they are also {\bf contextual}, i.e. that they can be given a physical sense only by embedding a microscopic system within a macroscopic measurement context.
\end{abstract}
\maketitle

\hspace{15pc}%
\begin{minipage}{25pc}
{\it Quantum mechanics occupies a very unusual place among physical theories: it contains classical mechanics as a limiting case, yet at the same time it requires this limiting case for its own formulation.} \\ 
\raggedleft  L.D. Landau \& E.M. Lifshitz, {\it Quantum Mechanics} (1965).
\vskip 2mm
{\it When you have eliminated all the impossible, then whatever remains, however improbable, must be the truth.}\\
\raggedleft  A. Conan Doyle, {\it The Case-Book of Sherlock Holmes} (1927).
\end{minipage}

\section{Prologue.}
 
 In 1923, Max Born gave a series of lectures at the University of G\"ottingen, from which he drew a monograph entitled ``The Mechanics of the Atom" \cite{Born}. 
The second edition of this book was published in 1927 in English, just after the founding articles on quantum mechanics, and he wrote in the Preface: 
\vskip 2mm

  {\it  ``Since the original appearance of this book in German, the mechanics of the atom has developed with a vehemence that could scarcely be foreseen. 
{\color{black}(... However,)} it seems to me that the time is not yet arrived when the new mechanics can be built up on its own foundations, without any connection with classical theory. 
It would be giving a wrong view of the historical development, and doing injustice to the genius of Niels Bohr, to represent matters as if the latest ideas were 
inherent in the nature of the problem, and to ignore the struggle for clear conceptions which has been going on for twenty-five years."}
\vskip 2mm

As we now approach the hundredth anniversary of these historic discoveries, can we say, as Max Born intended,  that {\it  the time is arrived when the new mechanics 
can be built up on its own foundations, without any connection with classical theory} ? The argument we want to develop here, and which -- as Max Born also wanted -- somehow 
does justice to the genius of Niels Bohr, is that this goal is in fact impossible to achieve.  Indeed, based on a century of theoretical and experimental progress, 
our conclusion is that classical physics and quantum physics can neither function nor even be conceived one without the other. 
Following the quotes above by Landau, Lifshitz and Conan Doyle this is a quite improbable situation indeed; 
nevertheless, 
it might well be the truth - we will see below why\footnote{The general level of the article corresponds to an undergraduate physics course. First year quantum mechanics students should follow easily the presentation, which deviates from textbooks more in the physical argumentation  than in the technical content. }.
 
\section{The Mysteries of Helium. }
 \vskip -2mm
The inability of classical physics to explain the structure of atoms, as well as the radiation they emit, is vividly illustrated in Max Born's book cited above.  By adding to the classical equations the heuristic 
recipe of angular momentum quantification proposed by Niels Bohr, it is possible to 
compute the radiation - the spectrum - emitted by hydrogen and some other (so-called hydrogenoid) atoms. But these calculations spectacularly  fail for helium, the simplest 
atom after hydrogen, with only two electrons - and Max Born's book demonstrates that this is not due to a miscalculation by early twentieth-century theorists. 
Mathematics is not magic, and correct calculations based on a physically false theory cannot provide results in accordance with experiments. 

But other experiments and ideas appeared during the 1920s: the Stern-Gerlach experiment proposed by Otto Stern and carried out with Walther Gerlach in 1922, the 
hypothesis by George Uhlenbeck and Samuel Goudsmit in 1925 of an intrinsic angular momentum of the electron, called spin a little later by Wolfgang Pauli; and finally the exclusion principle, 
by Pauli again, applied first to electrons, which are particles with half-integer spin, as detailed in standard textbooks \cite{TBQM1,TBQM2,TBQM3}.
Finally, in 1925-26, the situation changed, as Max Born recounts in the previous section, and the solution appeared, {\it with a vehemence that could scarcely be foreseen}. 
The two electrons in the helium atom are actually two spin 1/2 particles, strongly coupled by their electrostatic interaction, and subject to the Pauli exclusion principle. 
{\color{black} It was understood that their state} must be described mathematically by a so-called wave function, including their positions and spins, which must change sign (we say: be anti-symmetric) by permutation 
of the two electrons. 
There are thus two families of states: those that are symmetric for the spatial part of the wave function, depending on the electrons' positions, 
and antisymmetric for the spin; and those that are anti-symmetric for the spatial part and symmetric for the spin.
\vskip 1mm

Let's then focus on the lowest energy state 
of helium, called the ground state, which is symmetric for the 
spatial part and antisymmetric for the spin. The states of a spin 1/2 particle are traditionally denoted $|+\rangle$ and $|-\rangle$, corresponding to the measurement 
results $+\hbar/2$ and $-\hbar/2$ for the spin component along an arbitrary direction, often chosen along an $Oz$ axis. The symbol $\hbar = h/2\pi$ corresponds to a 
quantum of angular momentum, where $h {\color{black} \,\simeq \, 6.63 \; 10^{-34}J.s}$ is Planck's constant. For two electrons we can thus define 4 states, 
$|+ +\rangle$, $|+ -\rangle$, $|- +\rangle$, $|- -\rangle$, where the first $\pm$ relates to one of the electrons, and the second to the other. 
The corresponding values $\pm\hbar/2$ could in principle be obtained from independent measurements on the two electrons. 
None of these 4 states changes sign by permutation of the two electrons, so none of them is suitable for the ground state of helium. 
 \vskip 1mm

 {\color{black} Then, in order to describe this state, we have to}
consider that in the quantum formalism the symbol $| \psi \rangle$ for a quantum state, as introduced by Dirac \cite{TBQM1,TBQM2,TBQM3}, actually designates a vector in a 
mathematical sense: this means that linear combinations such as $(a |+\rangle + b |-\rangle)$ for one spin, or $(c |+ +\rangle + d |+ -\rangle + e |- +\rangle  + f |-  - \rangle)$ 
for two spins, are other possible states, the letters designating complex numbers. One can then see that the state 
$| s \rangle = (|+ -\rangle - |+ -\rangle)/\sqrt 2$ 
 {\color{black} provides a solution:} by swapping the two electrons we change $| s \rangle$ to $- | s \rangle$. This state $| s \rangle$ is called a `spin singlet',  we'll see why later, and 
the calculation, {\color{black} taking $| s \rangle$ as the spin part of the electrons' state,}  gives the correct values 
of the helium energy levels - a huge success for early quantum mechanics. 
\vskip 1mm

{\color{black} {\bf Now here is the true enigma,}   
that is the question which} has painfully plagued physicists for 100 years: if the electron pair is in the state $| s \rangle$, what is the value of the projection component 
of the spin of the first electron along the  
$Oz$ axis? And it unfortunately turns out that the formalism we have introduced above is completely incapable of answering this question. So let's be more concrete, 
and ask: in the $| s \rangle$ state, what value are we going to find by measuring the projection component of the spin of the first electron along the  
$Oz$ axis?  Then there is an answer: we can randomly find either $+\hbar/2$ or $-\hbar/2$, with probabilities both equal to 1/2.
\vskip 1mm

From here on, physicists tear themselves apart\footnote{Many, often conflicting,  interpretations on quantum mechanics are documented  in the Stanford Encyclopedia of Philosophy 
\cite{st1,st2,st3,st4,st5,st6,st7,st8}.}, and some say \cite{laloe}: but if we find $+\hbar/2$, it means that this projection was already 
worth $+ \hbar/2$ before we measured it, because it can't be otherwise, right? But the others answer: no, not at all, before the measurement the value of the projection was 
simply not defined - whatever it means. And still others say: in fact the measurement never happens, we have a branching of universes, one where the projection is $+\hbar/2$, 
and another where it is $-\hbar/2$. Still others claim that the result does not really exist, but that it is only a subjective bet made by the agent who made the measurement. 
And if we ask `but does the state $| s \rangle$ really exists', some will tell that this state may be ontic, epistemic, epi-ontic... which does not really help. 
\vskip 1mm
 
It is thus clear that a terrible slippage has occurred along the way, and that we must take a few steps back, and revisit the whole issue. {\color{black} 
In the line of our police investigation, the enigma is then: {\bf who or what caused the demise of the spin components of the individual electrons from our world? Did this really happen, and if yes, why and when? }}
This question amounts to no less than clarifying what exists, what is real, and what is objective in a quantum phenomenon. 
And this is not mere philosophy: at the moment where quantum technologies bloom and concepts of quantum physics must be assimilated by engineers, these points have to be clarified - or more 
stuff that was thought to be real  \cite{falcon} will keep on vanishing.
As a matter of facts, engineers, like enquirers,  do not work with metaphors that shed light on a deep but mysterious mathematical formalism.  They need predictable and repeatable facts on which to build intuition, and simple assessment rules because their action is at another complexity level, the one of building devices by getting many elementary systems to work together.
\section{One spin is almost fine, two spins are really weird. }
 \vspace{-1mm}
\subsection{Further investigations.}
\vspace{-1 mm}
Given the previous conundrum, it is appropriate to seek more expert advice, so let's call textbook quantum mechanics (TBQM) for more explanations \cite{TBQM1,TBQM2,TBQM3}. In the previous part we introduced the states  $|+\rangle$ and $|-\rangle$, corresponding to the results $+\hbar/2$ and $-\hbar/2$ of a measurement of the projection of the 
electron's spin in an $Oz$ direction.  Measuring the spin of an electron in an atom is not easy, but it can be done for atoms with a single `active' 
electron, this is the principle of the Stern and Gerlach experiment we have already talked about, initially carried out with silver atoms. 
%
The $Oz$ direction of the measurement corresponds to the direction of a magnetic field gradient that {\color{black} the atom experiences} in the device, and must be chosen beforehand; 
a measurement along $Oz$ does not tell what the result would be in another direction, $Ox$ or $Oy$ for example (see Fig. 1). Nevertheless, repeating the same measurement along $Oz$ will -- not 
surprisingly -- give the same result. So from now on we will add a subscript indicating the measurement direction, such as $|\pm_x \rangle$ for $Ox$, and $|\pm_z \rangle$ for $Oz$. 

 \begin{figure}[h]
\includegraphics[width=0.2 \columnwidth]{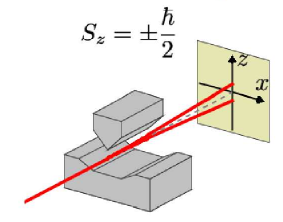}
\includegraphics[width=0.2 \columnwidth]{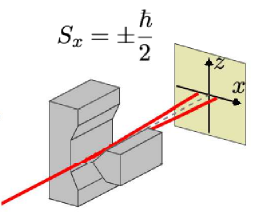}
\includegraphics[width=0.4 \columnwidth]{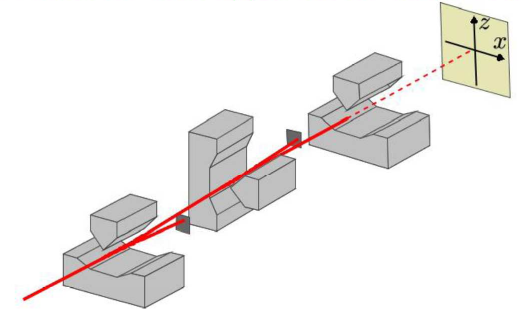}
\caption{\label{SG} 
Stern Gerlach magnets measuring the spin components along $Oz$ (left), along $Ox$ (center), 
and along $Oz / Ox / Oz$ (right). In this last case the results $-\hbar/2$ are blocked by the dark square screens. 
\textcopyright~Manuel Joffre, Ecole Polytechnique. }
\end{figure}

We can then try to chain several measurements: first along $Oz$, which gives let's say $+\hbar/2$, then {\color{black} along} $Ox$, then {\color{black} along} $Oz$ again. 
Theory and experiment say that the second measurement, along $Ox$, gives a random result $\pm \hbar/2$, with probabilities of 1/2. 
Why not, so let's assume that we find $+\hbar/2$, and let's go back to the projection {\color{black} component} along $Oz$ that we knew before; but if we make the measurement 
again along $Oz$, we find a random result $\pm \hbar/2$, with probabilities of 1/2: {\bf everything happens as if measuring along $Ox$ had erased the previously known result along $Oz$~!} {\color{black} So~already with a single spin 1/2 particle, a strange  vanishing of some properties can be observed - maybe a hint of what is going on with the more tricky two-spin situation.}
\vspace{2 mm}

Another very strange fact is this: suppose Alice randomly selects a  particle in any one of the 4 states $|\pm_x \rangle$ or $|\pm_z\rangle$, and sends it to Bob without any further indication.  
Bob is then unable to determine the state of the particle without error!  Indeed, if he chooses (by chance) the same measurement axis as the one selected by Alice, he will find the 
right result, the one she has sent, but if he chooses another axis, he will find a random result, uncorrelated with what Alice sent.  This impossibility, which also applies to any eavesdropper,  is at the basis of the technique called quantum 
cryptography, or more precisely prepare-and-measure quantum key distribution. 
But then how it is possible for Bob to get any useful information? This can be done by Alice and Bob publicly exchanging their choice of axis {\bf after} Bob has received the particle, and keeping only the results where they made the same choice; the revealed information is then useless for a possible eavesdropper, since the particle is no more there  \cite{QKD}.
\vspace{2 mm}

The appearance of these random results raises several questions:
\vspace{1 mm}

- Is it possible to find, in one way or another, {\color{black} situations where a measurement provides a  deterministic result,  i.e. that can be predicted with certainty? This seems mandatory in order to attribute a 
physical reality to the objects we consider, 
and in the framework of our inquiry, we need actual and objective facts, not ghostly ones.}

- If Bob can't fully identify the state of the particle (the one known and sent by Alice) by making measurements on that particle, is it legitimate to say that $|\psi \rangle$ is the `complete state of the particle'? 
\vspace{2 mm}

Usual TBQM attributes a `complete quantum state' to a particle, so its answer to the second question above is affirmative, but it tells  also that 
despite this completeness, Bob's measurement result is generally not deterministic. Then, moving on to two particles, the contradictions accumulate, 
culminating in the strange situation mentioned at the end of the previous section. {\color{black} Therefore, unfortunately,  one has to conclude that the expert advice from TBQM did not really help, but rather increased the confusion. }
So we will now look for different answers to these questions, by telling that the quantum 
state may not be so complete after all –- and show that we can thereby make one step towards elucidating this two-spin enigma, in helium and everywhere else. 
 
\subsection{Some clues and first answers to the questions.}
In the rest of this article, we will reason about idealized experiments, in the sense that they are possible without contradicting anything we know about quantum mechanics (QM), 
but can be very difficult to perform in practice. Such reasoning has been used since the early days of QM, under the name of thought experiments, or {\color{black} Gedankenexperiments}.  
In fact, it turns out that nowadays the progress in the mastery of quantum systems is such that a very large number of these thought experiments have been carried out, with such a 
good degree of approximation that they can be considered to experimentally validate the reasoning initially carried out in principle. 
 \\

This is particularly the case for what are now called `Quantum Non-Demolition measurements', or QND measurements, in which the quantum state of a particle is identified, 
while leaving that particle in the observed state afterwards \cite{QND}. These measurements are sometimes referred to as von Neumann's ideal measurements, because John von Neumann used them 
extensively from a theoretical point of view \cite{JvN}; but it is now possible to consider them as well-established experimental facts \cite{qnd_pg,sh_jmr,qnd_wineland}. 
They have the considerable advantage of giving a clearly positive answer to our first question, i.e. whether it is possible to obtain, in one way or another, deterministic results. 
The answer is yes, because it is enough to repeat a QND measurement on the same particle or system, to find again the same result with certainty.  Let us insist on the essential 
point that it is not only a question of chaining together different measurements on the same system, which leads to random results as we have written before: it is a question of 
{\bf repeating the same measurement on the same system.} This certainty may not be as general as we might have liked, because it is destroyed by making a different measurement 
as said before, but it is consistent with what we learn in QM textbooks and observe in the lab, so it will allow us to move forward. 
 \vskip 2mm

The answer to the second question, `if Bob can't find the state of the particle by measuring that particle, is it legitimate to say that it is indeed the state of the particle?' 
is going to take us further off the beaten track.  We've already seen that if Alice gives Bob the particle without indicating its state, Bob can't get back to that state 
with certainty. On the other hand, if Alice gives Bob the particle without indicating its state, but also indicating the direction of measurement she used, $Ox$ for example, 
then Bob can identify with certainty which state, $|+_x\rangle$ or $|-_x\rangle$, was {\color{black}prepared}  by Alice. We thus find the same certainty as before, provided again 
that the correct measurement is repeated, i.e. that Alice's and Bob's axis agree. Introducing a bit of terminology, we'll call the set of classical parameters defining Bob's 
action a  `measurement context', or more briefly, a {\bf context} denoted $\mathcal{C}$ and 
materialized by a macroscopic classical device. 
We can thus say that the physical object that 
owns a certain and reproducible measurement result is not the system (the particle) alone, 
but {\bf the system within a context. } 
{\color{black} In this framework} the possible results of a specified measurement, attributed to a system within a context, are %
called {\bf modalities}~\cite{CSM1}. 
 \vskip 2mm

If Bob has the quantum particle and knows the 
context used by Alice to prepare the particle, then he can identify an associated modality. 
For a single spin the context is defined by the direction of the projection, $Oz$ for example, and there are only two mutually exclusive results, $\pm \hbar/2$. 
A possible modality is then $+\hbar/2$ along $Oz$, which corresponds to the vector $|+_z\rangle$ introduced earlier. For a single spin  it seems that there is no difference 
between a state and a modality, but we will see later that the distinction between the usual (textbook) state vector and the modality becomes essential for larger systems, for example for two spins.  
 \vskip 2mm

To summarize this section, we have found some determinism in the behavior of the physical objects used in QM, if we accept that they are in fact (quantum) systems within (classical) contexts. 
{\color{black} Then the very existence of the victim of the demise - the spin component of the electron - may become uncertain indeed, because it is no more a property belonging to the electron alone. Clearly the modality may change if we change  the context - and a new modality will show up in a new context, and will be then certain and repeatable.
But before exploiting this major result in our inquiry,} let's go back once again to helium, and more precisely to the other family of helium states,  
 those that are anti-symmetric for the spatial part and symmetric for the spin - that is, their spin state does not change its sign by permutation of the two electrons.
 
\subsection{The solution to the enigma – or not? } 
 
The 4 possible states $|+_z +_z\rangle$, $|+_z -_z\rangle$, $|-_z +_z\rangle$, $|-_z -_z\rangle$ for the two spins correspond to separate measurements made on the two electrons, 
and do not all fit with the symmetry or anti-symmetry condition imposed by the Pauli exclusion principle. We have already seen that the vector 
$| s \rangle = (| +_z -_z \rangle - | -_z +_z \rangle)/\sqrt 2$, the singlet state, is antisymmetric. But what are the symmetric states? 
One can see that $|+_z +_z\rangle$ and $|-_z -_z \rangle$ are suitable, as well as the state $| r \rangle = (| + _z -_z\rangle + | -_z +_z \rangle)/\sqrt 2$: we therefore have 
three symmetric states, globally called triplet states. 
 \vskip 2mm

We have already said that the symbols $| \psi \rangle$ are vectors, now it must be specified that the four vectors $| \pm_z \pm_z \rangle$ are in fact orthogonal, and correspond 
to measurement results, modalities 
as defined above, which are mutually exclusive: only one of these 
four results can be obtained when measuring the components along $Oz$ of the spin of each electron.  Let us then consider the four states $|+_z +_z\rangle$, $| s \rangle$, 
$| r \rangle$, $|-_z -_z\rangle$.  An elementary calculation shows that they are also orthogonal, and thus also describe mutually exclusive measurement results, but what is 
the associated measurement? A standard algebraic calculation in QM shows that this is the measurement of the quantities $\vec{S}^2$ and $S_z$, where $\vec{S}$ is  the total spin of the two electrons 
{\color{black}$\vec{S} = \vec{S}_1+ \vec{S}_2 $}; it is associated with the two quantum numbers $S = 0$ and $S = 1$ 
(see Fig. 2).  For $S = 0$, a measurement of the 
total spin along $Oz$ (or an arbitrary axis) gives the value 0: this is in fact the state $| s \rangle$. For $S = 1$, a measurement of the total spin along $Oz$ gives 
the values $+\hbar$ for the state $| +_z +_z \rangle$, 0 for the state $| r \rangle$, and $–\hbar$ for the state $| -_z -_z \rangle$. 
So we again have 4 mutually exclusive results, which are noted $| S=0, m=0\rangle$, $| S=1, m=1\rangle$, $| S=1, m=0\rangle$, $| S=1, m=-1\rangle$, where $m$ is the result 
$m \hbar$ of the measurement. We note also that we can write $| S=1, m=1\rangle = |+_z +_z\rangle$ and  $| S=1, m=-1\rangle = |-_z -_z\rangle$: these are therefore the same 
`state vectors', but not the same modalities, since these results appear in completely different measurement contexts, corresponding either to measurements on the separate spins, 
or to a global measurement on the total spin. This distinction is quite important \cite{extra}, and we will return to it later; see Fig. 2 for a summary. 
\\
 \begin{figure}[h]
\includegraphics[width= 0.95 \columnwidth]{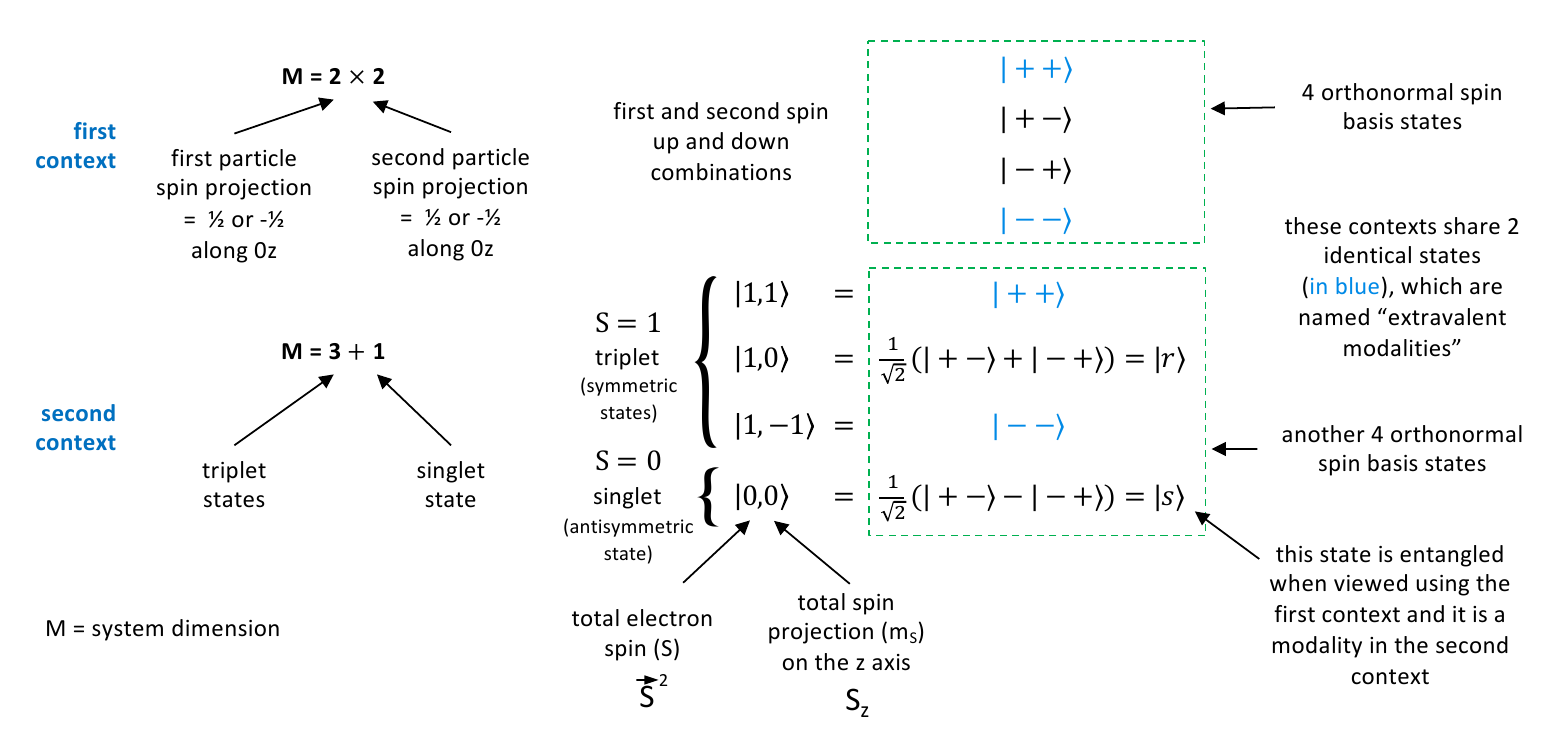}
\caption{\label{He} 
Various quantum states relevant for two electrons. All $| \pm \rangle$ mean $|\pm_z \rangle$. The total spin
$\vec{S}^2$ is the sum of the squares of the three spin projections $(S_x^2 +S_y^2+S_z^2)$ 
and it takes the values $S(S+1) \hbar^2$. {\color{black} In this figure $M$ is the total number of possible, mutually exclusive measurement outcomes -- or modalities --  in each context.}
}
\end{figure}

Though the inquiry is clearly  making progress,  we will conclude this part with a new conundrum, by again considering the states/modalities $| s \rangle$ and $| r \rangle$, perfectly defined for a measurement
of the total spin, and associated with well-defined energy levels of the helium atom. As we have already seen, the question `what is the value of the component of 
the spin along $Oz$ for only one of the two electrons' {\it has no answer} within the framework of the formalism of usual QM we have introduced. 
We can reconcile ourselves with this statement by saying: of course there is no answer, but this is not really shocking since we are talking about two electrons within the same atom, 
very strongly coupled by electrostatic interaction, so it is not very surprising that they adopt a global behavior. In addition, there is no clear way to measure each electron's  spin within the helium atom. 
\\

Yet the ever-doubting physicist will ask the question: ok, but can't we have the two electrons, or any two other spin 1/2 particles,  
very far apart, and still in the state $| s \rangle$ or in the state $| r \rangle$? This can be done indeed in suitable experiments, as detailed in the Appendix. What, then, 
is the meaning of the statement that the component of its spin along $Oz$ has no value, or a completely random value? And what happens if we compare the results of separate 
measurements on the two spins? Then we have to dive into an even deeper mystery,
that of quantum entanglement, Verschr\"ankung 
in German, with two famous papers published in 1935: `Can Quantum-Mechanical Description of Physical Reality Be Considered Complete?', by Einstein, Podolsky and Rosen (EPR) \cite{EPR}, 
and Schr\"odinger's almost desperate article, though resignated to the quantum weirdness, `Die gegenw\"artige Situation in der Quantenmechanik (The present situation in quantum mechanics)' \cite{cat}. 
{\color{black} The cause of desperation of these great physicists was the following: the spin components of the two remote electrons, that were supposed to have ``no value", turn out to be strongly correlated! But how can it be, if their values are undetermined? 
Did QM fall from Charybdis to Scylla? It is this new extended enigma  that will now have to be solved. }
 
{\color{black} \section{ Predictive incompleteness of $| \psi \rangle$ with no context.  }}
\subsection{The Bell perspective.}
The problems raised by Einstein and his colleagues in 1935 addressed the almost philosophical foundations of QM but had no direct experimental implications. 
The situation changed in 1964 when John Bell proposed inequalities, deduced from hypotheses very close to those discussed by EPR, and which could be tested experimentally 
on entangled pairs of particles \cite{Bell}. This stimulated a long series of experiments, starting in 1972, with highlights in 1982 and 2015 \cite{Aspect}, that fortunately 
concluded with the Nobel Prize awarded in 2022 to Alain Aspect, John Clauser and Anton Zeilinger, `for experiments with entangled photons, establishing the violation of Bell's 
inequalities and pioneering quantum information science'.  We are not going to tell this story here, since it has been the subject of many articles and books, but rather consider 
the question: {\bf what is ultimately the content of Bell's hypotheses, leading to these inequalities, and now dismissed by observations?} Many possible answers have been proposed, 
but we will now  develop one that is in line with the content of the previous sections. 
 \vskip 2mm

A possible understanding of Bell's perspective is by telling that regardless of the  
limitations of actual measurements, the underlying 
reality should have a definite configuration, described by some local hidden variables (LHV),
that fully specifies everything that can/will happen. 
The existence of a physical mechanism producing an effect from LHV
as a cause is very generally true in classical physics, 
and it is also built in the use of classical probability theory. It means that even if the 
values of the LHV
are not known, they are supposed to exist as the `hidden cause'. Such an explanatory mechanism is generally called the {\it completeness of inference} {\color{black}(or predictive completeness), given the  hidden parameters \cite{inference}. 
Let us emphasize that this predictive completeness}
does not mean fully knowable determinism but ontological determinism {\color{black} i.e. the existence of a fundamental 
underlying predictable mechanism}, as expressed by Laplace's demon or by Einstein's famous quote ``God does not play dice''. What is at stake is thus the possibility 
to specify everything that will happen, in a classically deterministic or {\color{black} at least in a} probabilistic theory, which is also assumed by Bell to be local. 
Equivalently, one may tell  that randomness has an ``ignorance interpretation": like in tossing a coin, the result is random, but in principle knowing all the initial parameters and the laws of dynamics should allow one to predict the result with certainty. 
 \vskip 2mm

Now, without further ado or mathematics, it is clear that the usual quantum state vector, as introduced in the previous section for describing the spin systems and generically 
denoted by  $\vert \psi\rangle$, does not fit in this framework. Specifying everything that may happen {\color{black}{\it can only be done once the measurement context is specified}}. 
More precisely, this measurement context is required, in addition to $|\psi\rangle$, in order to define a standard probability distribution over a set of mutually exclusive events. 
Specifying all these possible events (i.e. the possible measurement results) is equivalent to giving the context, and adding $|\psi\rangle$ allows one to calculate their respective probabilities to occur. {\color{black} In this sense, we can say that the quantum state alone is {\it predictively incomplete}. And since complete predictiveness is obtained by specifying the context, we have to acknowledge that what QM allows are {\it contextual inferences} \cite{inference}}.
 \vskip 2mm

Though this just tells how quantum mechanics works, it may appear shocking to both a classical physicist and a (naive) quantum physicist. A classical physicist has a hidden cause 
of randomness and correlations (Bell's LHV)  
deeply rooted in his mind, so telling that there is no such variables in quantum physics is hard to accept. 
On the other side, a quantum physicist has deeply rooted in his mind that $|\psi\rangle$ is complete, in the sense that there is no underlying hidden cause to be sought - as it 
was taught by Niels Bohr, and as it is still true. But telling that quantum state vector $|\psi\rangle$ is incomplete as long as the context has not been specified, though it 
is operationally obvious in quantum mechanics, and does not contradict Bohr's message, may be quite hard to accept. Why? Because we would have then to give up the idea that 
$|\psi\rangle$ fully specifies a physical object, which is the quantum system under study, and to admit that {\bf the actual physical object has to be the quantum system, 
within a classical context.} This appears so shocking that the quantum physicist may be tempted to consider 
 {\color{black} instead} `non-local influences' or other weird explanations, in order to violate 
Bell's inequalities; but this leads to the contradictory points of views presented before, and it is actually not needed if predictive incompleteness is acknowledged \cite{inference}. 
{\color{black} \subsection{Mutual consistency of the classical and quantum descriptions of the physical world.  }}
The idea that a context is needed to get a complete description of what can be observed on a quantum system has often been confused with statements about the subjectivity of the 
quantum state, or about the role of the consciousness of the observer \cite{laloe}. There is no need for that either, and a physical object remains perfectly objective,  but still it is a system within a context. 
This requires however to recognize that there is no such thing as absolute objectivity of predictions on the system alone; 
this departs 
from the usual, classical, apprehension of the universe, taken as a whole, with all its logical and ontological necessity as Newton and Einstein conceived it. The quantum 
requirement for a contextual description of physical objects is related to  many issues in quantum physics, and it has been spelled out initially in the article `Contextual objectivity: a realistic interpretation of quantum mechanics' 
(2001) \cite{contextual}, which may provide us with a more suitable, non-classical view about the world we live in \cite{nayla}. 
  \vskip 2mm

There is clearly a price to pay for the idea that there are well defined systems and contexts: if this is the case, there must be a physical separation between them, 
usually known as the ``Heisenberg cut",  {\color{black} where the applicable laws change from quantum to classical}. This creates a dualist view of the physical world, that does not seem satisfactory, and there have been many attempts to find ways 
by which the classical world should `emerge' from the quantum one -- if any. This was also the content of Max Born's initial request,  to build up `the new mechanics on 
its own foundations, without any connection with classical theory'.
But it can also be told, as Bohr and Landau did\footnote{According to Landau \& Lifshitz \cite{TBQM1}, pages 2 and 3, ``It is clear that, for a system composed only of quantum objects, it would be entirely impossible to construct any logically independent mechanics (...) Thus quantum mechanics occupies a very unusual place among physical theories: it contains classical mechanics as a limiting case, yet at the same time it requires this limiting case for its own formulation." For a possible explanation of this strange loop, see \cite{MP2}, Figure 3. }, that there is no way to formulate quantum mechanics without referring to classical concepts -- here come again 
the contexts, as discussed above. This leads to the idea that {\bf classical and quantum descriptions are both required from the beginning and simultaneously for mutual 
consistency}  \cite{myst,completing}. Actually, classical without quantum is unable to explain the structure of atoms as seen above, but quantum without 
classical is unable to tell clearly what a quantum state describes, and when extrapolated to the macroscopic world it gets lost in bizarre predictions clashing with empirical 
evidence. 
  \vskip 2mm

An apparent difficulty for unifying classical and quantum physics is that their mathematical frameworks seem incompatible -- but this is not correct, and a unified formalism, 
using operator algebras, is actually possible and was already suggested by John von Neumann in 1939. Though this extension of the standard quantum formalism is mathematically 
nontrivial, it definitively deserves a closer look by quantum physicists -- after being placed into the appropriate ontological framework introduced above. 
This has been discussed in several recent articles \cite{MP1,MP2,MP3}.
  \vskip 5mm

{\color{black} \subsection{Recovering real properties: modalities vs state vectors.}}
An important step in our discussion has been to introduce a specific word representing a physical property or set of physical properties, attributed to a system within a context - we call it a modality. Clearly a modality is different from a usual quantum state vector $|\psi\rangle$, which does not specify the context. It is important to note that for a single spin $1/2$ particle, just giving a quantum state vector like $|+_z\rangle$ actually specifies the context $Oz$, and there is only one other mutually exclusive modality, that is $|-_z\rangle$. But for larger values of the spin, or for more particles, there are more than two mutually exclusive modalities in a given context. In that case, a given quantum state vector may appear in infinitely many contexts, as it was the case (for two different contexts) with $| +_z +_z \rangle = | S=1, m=1 \rangle$ in the example given above. 
  \vskip 2mm

One should also note that the quantum state vectors $| \pm_z \rangle, | \pm_x \rangle$ are related by 
$| +_x \rangle = (| +_z \rangle + | -_z \rangle)/\sqrt{2}$ and $| -_x \rangle = (| +_z \rangle - | -_z \rangle)/\sqrt{2}$: 
the state vectors along $Ox$ are linear superpositions of the state vectors along $Oz$. It is often told that `quantum superposition are like being in the two states 
$| +_z \rangle$ and $| -_z \rangle$ at the same time', which clearly makes no sense: the modalities associated with $| \pm_x \rangle$ are certainties in the $Ox$ 
context, mathematically expressed as linear combinations of state vectors $| \pm_z \rangle$ that are certainties in the $Oz$ context. It is also useful to keep in 
mind that the state $| s \rangle$ appears as entangled in the context where the two spins are described and measured separately, but not in the context based on the 
total spin with $S=1$ or $S=0$. Entanglement is thus related to the choice of a description and measurement context (technically, a tensor product structure) for the 
system of interest. Obviously, the description using $| \pm, \pm \rangle$ appears more `natural' when the two spins are spatially separated. 
  \vskip 2mm

Putting everything together, the theoretical tools used so far consist in associating orthogonal vectors to mutually exclusive modalities (in the same context), 
and the same vector to mutually certain modalities (in different contexts). Introducing some vocabulary, the non-obvious physical phenomena that certainty can be transferred 
between contexts is called extracontextuality \cite{extra}.  Then the property of  `being mutually certain' defines an equivalence relation between modalities, which is 
called extravalence  - think again of the relation $| +_z +_z \rangle = | S=1, m=1 \rangle$. 
  \vskip 2mm

We come thus to the conclusion that {\bf the quantum state vector $|\psi\rangle$ is a mathematical object, associated with an extravalence class of modalities.} This establishes 
a clear separation between mathematical objects like $|\psi\rangle$, and physical objects that are systems within contexts, and carry well-defined and measurable properties 
that correspond to modalities. These sentences require some thinking to be fully appreciated, but they do refer to the ontology of physical 
objects (systems and contexts) and their properties (modalities) – within the non-classical but realistic framework provided by contextual objectivity. 
  \vskip 2mm

To complete the picture we need a fundamental postulate, called contextual quantization, which tells that {\bf the maximum number $M$ of mutually exclusive modalities is a 
property of the quantum system, and is the same in any relevant context.} We have already seen an example of this behavior with the two electron spins, where $M$ = 4 
for several possible contexts – this is actually the case for  
any such context. Given this postulate, it can be shown that the relation 
between modalities in different contexts can only be probabilistic \cite{random}. The fixed value of $M$ also allows one to associate the system with a $M$-dimensional 
vector space – actually a Hilbert space \cite{Gleason,Uhlhorn}.  In this picture there is no trouble in using the standard projection postulate of textbook QM, that 
appears simply as an update of the probability distribution associated with $|\psi\rangle$, for a given context. 
\vskip 3mm

The crucial point is that this probability distribution is non-classical and contextual, and gets a  {\color{black} computable} physical meaning only once the actual context has been specified \cite{random}. 
For the two-spin  enigma, the violation of Bell's inequalities is attributed to the predictive incompleteness of $|\psi\rangle$, and it does not 
require any non-local influence, but rather a contextual inference,
as it is  explained in details in \cite{inference}. So there is no `hidden variables' to be looked for, and the context, completing $|\psi\rangle$, specifies the ``very conditions 
which define the possible types of predictions regarding the future behavior of the system'', as stated by Bohr \cite{Bohr}. Mixing up probabilities calculated in 
different contexts is counterfactual, and leads to contradictions. There is no doubt that these explanations are quite remote from classical physics, 
but again they make full sense in the framework of contextual objectivity. 
\vskip 6mm

\section{Epilogue.}
 
Since the purpose of this paper is to illustrate and discuss quantum ideas  using  two spin $1/2$ particles, 
we did not reconstruct the full quantum formalism yet, and explaining basic tools like unitary transforms and Born's rule are still missing. But  there is not much 
choice left, because the framework defined above fits with the hypotheses of powerful mathematical theorems, establishing the need for unitary transformations 
(Gerhard Uhlhorn's theorem in 1963) and for Born's rule (Andrew Gleason's theorem in 1957).  These arguments have been published elsewhere \cite{Gleason,Uhlhorn} and will not be reproduced here.
Also, systems and contexts should be part of a unified mathematical formalism, going beyond standard textbook QM. Operators algebra, introduced by Murray and 
von Neumann \cite{rings,infinite} and considerably developed since then, can provide such a formalism and give a mathematical status to the Heisenberg cut, as it has been shown in \cite{MP1,MP2,MP3}.
 \vskip 2mm

So finally, what should we tell to the layman, 
to bring our investigation to a successful conclusion, 
and to replace misleading statements such as ``a quantum superposition is like being in two states at the same time'', 
or ``quantum entanglement is like an instantaneous action at a distance''? It could be: 
 \vskip 2mm

{\it The physical properties of microscopic systems are {\bf quantized}, as initially shown by Planck and Einstein, and they are also {\bf contextual}; this means that they 
can be given a physical sense only by embedding a microscopic system within a macroscopic context that specifies how the system is observed,  as proposed by Bohr. Such a behavior is quite remote from the  
non-quantized  and non-contextual perspective of classical physics,  and it requires developing a specific contextual  probabilistic theory, which is basically what quantum mechanics has been doing. }
 \vskip 2mm
 
 To conclude, let's emphasize that QM follows from the { \bf conjunction of quantization and contextuality}; these two properties taken separately may have some classical analogue, and therefore they don't have the same constraining power than when being combined. 

\vskip 3mm

\section{Appendix: Another view on two spin 1/2 particles.}

In the above text we use the example of the helium atom to introduce the four states 
$|+_z +_z\rangle$, $| r \rangle$, $| s \rangle$, $|-_z -_z\rangle$
that can also be written
$| S=1, m=1\rangle$, $| S=1, m=0\rangle$, $| S=0, m=0\rangle$, $| S=1, m=-1\rangle$.
In the helium atom these states are  imposed by the Pauli exclusion principle, and 
can be identified because they correspond to specific energy levels.
On the other hand, the context corresponding to the separated states of the two electrons 
 $|+_z +_z\rangle$, $|+_z -_z\rangle$, $|-_z +_z\rangle$, $|-_z -_z\rangle$
 is not accessible within the helium atom, since this would require to 
 have individual access to  the electrons {\color{black}  by taking them away far from the atom and thus changing their energy state}. 
 \\
 
 So it would be nice to exhibit another system, where all these states can be manipulated and measured. 
 This has been proposed by photo-dissociating a Hg$_2$ molecule in a singlet state, and carrying out spin measurements on the two remote fragments~\cite{fry}. But this turned out to be a quite difficult experiment, and another more feasible approach is to consider that a spin 1/2 particle is actually a quantum bit, a qubit in a quantum computer, 
 and by identifying $|+_z \rangle$ with the logical state $|1\rangle$, and $|-_z \rangle$ with the logical state $|0\rangle$. 
 \\
 
 The question is then to move from the four states (`uncoupled basis') 
 $|1,1\rangle$, $|1,0\rangle$, $|0,1\rangle$, $|0,0\rangle$, usually called the computational basis, 
 to the four states   (`coupled basis') 
 $|1,1\rangle$, $(|1,0\rangle+|0,1\rangle)/\sqrt{2}$, $(-|1,0\rangle+|0,1\rangle)/\sqrt{2}$, $|0,0\rangle$.
This can be done by the following $4 \times 4$ unitary matrix:
 \begin{equation*}
G = \left( {\begin{array}{*{20}{c}} 
1&0&0&0\\
0&1/\sqrt{2}&-1/\sqrt{2}&0\\
0&1/\sqrt{2}&1/\sqrt{2}&0\\
0&0&0&1
\end{array}} \right)
\end{equation*}
which can be implemented by a sequence of quantum gates, as shown on Fig. 3. 
 This shows explicitly that, as written in the text, two qubits can be set or measured in the state $| s \rangle$ or in the state $| r \rangle$  outside the helium atom.   
\begin{figure}[h]
\begin{center}
\includegraphics[width=0.75 \columnwidth]{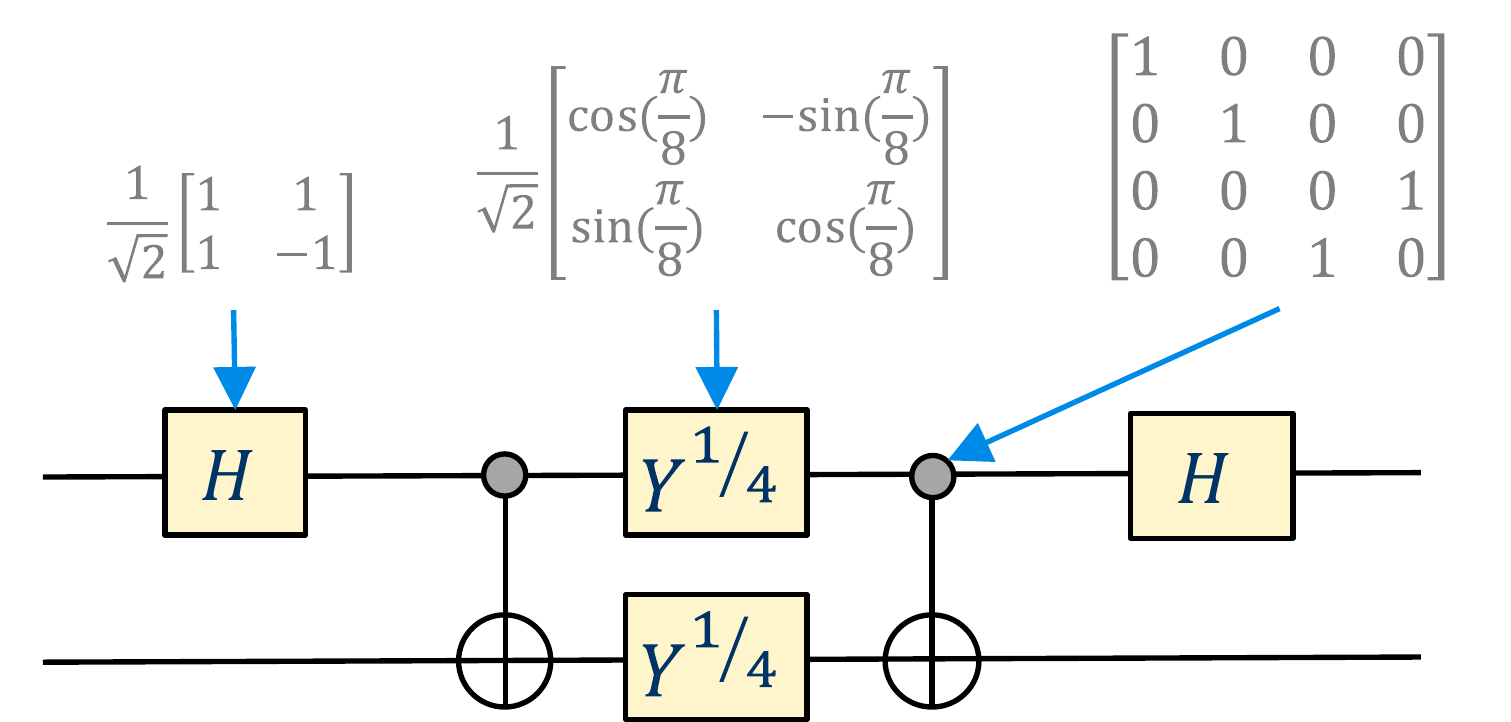}
\end{center}
\caption{\label{gates} 
Sequence of one- and two-qubit quantum gates and their matrix unitaries acting on two qubits, 
to move between the coupled and uncoupled basis. The first and last gates are Hadamard gates, the second and forth Controlled-NOT gates, and the middle ones are specific non-Clifford one-qubit gates. }
\end{figure}

{\color{black} An arbitrary quantum state can thus be measured  either the computational basis
 $\{ |1,1\rangle$, $|1,0\rangle$, $|0,1\rangle$, $|0,0\rangle \}$, or in the coupled basis 
 $\{ |1,1\rangle$, $(|1,0\rangle+|0,1\rangle)/\sqrt{2}$, $(-|1,0\rangle+|0,1\rangle)/\sqrt{2}$, $|0,0\rangle \}$.
 More precisely, the non-trivial measurement in the coupled basis can be realized by applying $G^\dagger$ to the two qubits, carrying out a QND measurement in the  uncoupled (computational) basis, which is in principle easy, and then applying $G$ to recover the measured state in the coupled basis. }
Therefore, such a given sequence of gates, building up a unitary transformation, implements a controlled change of context, since not only one but a whole set of mutually orthogonal states are mapped onto another set of mutually orthogonal states. This has the advantage that modalities can be defined up to a unitary transform, in the case where the appropriate context cannot be implemented. Though not so easy to implement in practice, this two-way procedure ($G^\dagger$ / QND / $G$) is useful in quantum computing, and shows that thinking about physical modalities, associated with a context, may be more enlightening than just thinking about mathematical state vectors, which represent extravalence classes of modalities occurring in different contexts. 


\vspace{1 cm}

\noindent {\bf Acknowledgments: }
PG thanks Franck Lalo\"e, Roger Balian, and Karl Svozil for many interesting and useful discussions, and Manuel Joffre for permission to use Fig.~1 from his Quantum Mechanics course at Ecole Polytechnique. MvdB thanks the Thales engineers for their candid but actually extremely relevant questions. The authors thank Michel Kurek and Camilia Ben Messaouda for proof reading and comments on  the paper.


\begin{thebibliography}{99}

 \bibitem{Born} Born, M. (1927) {\it The Mechanics of the Atom}, G. Bell and Sons, London.
 
 \bibitem{TBQM1} 
Landau,  L.D., Lifshitz, E.M. (1965)  {\it Quantum Mechanics}, available online: 
https://archive.org/details/ost-physics-landaulifshitz-quantummechanics ; 
 \bibitem{TBQM2} Feynman, R. (1963) {\it The Feynman Lectures on Physics, Volume III: Quantum Mechanics},  
available online:  https://www.feynmanlectures.caltech.edu/~;
 \bibitem{TBQM3}  Cohen-Tannoudji, C., Diu, B., Lalo\"e, F.  ( 2019) {\it  Quantum Mechanics} (3 Volumes),
 John Wiley \& Sons. 
 
 \bibitem{laloe} Lalo\"e, F.  (2012) {\it Do We Really Understand Quantum Mechanics?}, Cambridge University Press.
 
\bibitem{st1} Faye, J. (2024) 
``Copenhagen Interpretation of Quantum Mechanics", \\
https://plato.stanford.edu/entries/qm-copenhagen/
 
\bibitem{st2} Bacciagaluppi, G. (2020) 
``The Role of Decoherence in Quantum Mechanics", \\
https://plato.stanford.edu/entries/qm-decoherence/

\bibitem{st3} Goldstein, S. (2021) 
``Bohmian Mechanics", \\
https://plato.stanford.edu/entries/qm-bohm/

\bibitem{st4} Ghirardi, G \& Bassi, A. (2020) 
``Collapse Theories", \\
https://plato.stanford.edu/entries/qm-collapse/

\bibitem{st5} Barrett, J. (2023) 
``Everettian Quantum Mechanics", \\
https://plato.stanford.edu/entries/qm-everett/

\bibitem{st6} Vaidman, L.  (2021) 
``Many-Worlds Interpretation of Quantum Mechanics", \\
https://plato.stanford.edu/entries/qm-manyworlds/

\bibitem{st7} Lombardi, O. \& Dieks, D.  (2021)  
``Modal Interpretations of Quantum Mechanics", \\
https://plato.stanford.edu/entries/qm-modal/

\bibitem{st8} Laudisa, F. \& Rovelli, C. (2019)  
``Relational Quantum Mechanics", \\
https://plato.stanford.edu/entries/qm-relational/

\bibitem{st9} Healey, R. (2022) 
``Quantum-Bayesian \& Pragmatist Views of Quantum Theory", \\
https://plato.stanford.edu/entries/quantum-bayesian/

 
 \bibitem{QKD} Bennett, C. H.  and Brassard G. (1984) ``Quantum cryptography: Public key distribution and coin tossing". In Proceedings of IEEE International Conference on Computers, Systems and Signal Processing, volume 175, page 8. New York.
 
 \bibitem{falcon} Bogart, H. (1941) ``The stuff that dreams are made of'', 
 in ‘The Maltese Falcon’, probably inspired from W. Shakespeare, ‘The Tempest’, Act IV, Scene 1.
 https://www.youtube.com/watch?v=qK9hLJsCWmg
 
  \bibitem{QND} Braginsky, V.B. and Khalili, F.Y.  (1996) ``Quantum nondemolition measurements: the route from toys to tools",   Rev. Mod. Phys. 68, 1. 
  
     \bibitem{JvN} von Neumann, J. (1955) {\it Mathematische Grundlagen der Quantenmechanik}, Springer, 1932; English translation {\it Mathematical Foundations of Quantum Mechanics}, Princeton University Press.
  
  \bibitem{qnd_pg} Grangier, P., Levenson, J.A., Poizat, J.P.  (1998) ``Quantum Non-Demolition Measurements in Optics’’, Nature 396, 537.

\bibitem{sh_jmr} Haroche, S. and Raimond, J.M. (2006) {\it Exploring the Quantum},  Oxford University Press, Oxford.

\bibitem{qnd_wineland} Hume, D.B., Rosenband,  T., Wineland, D.J. (2007) ``High-Fidelity Adaptive Qubit Detection through Repetitive Quantum Nondemolition Measurements’’, Phys. Rev. Lett. {\bf 99}, 120502.
  
   
    \bibitem{CSM1} Auff\`eves, A. and Grangier, P. (2016) ``Contexts, Systems and Modalities: a new ontology for quantum mechanics'', Found. Phys. 46, 121  [arXiv:1409.2120].
 
  \bibitem{extra} Auff\`eves, A. and Grangier, P. (2018)  ``Extracontextuality and extravalence in quantum mechanics",  Phil. Trans. R. Soc. A 376, 20170311 [arXiv:1801.01398].
  
 \bibitem{EPR}  A. Einstein, B. Podolsky and N. Rosen  (1935) ``Can Quantum-Mechanical Description of Physical Reality Be Considered Complete?", Phys. Rev. 47, 777.

 \bibitem{cat} Schr\"odinger,  E.W. (1935) ``Die gegenwärtige Situation in der Quantenmechanik (The present situation in quantum mechanics)", Naturwissenschaften. 23, 807. 
 
 \bibitem{Bell}  Bell,  J.S.  (1964) ``On the Einstein Podolsky Rosen Paradox", Physics {\bf 1},195–200.
 
 \bibitem{Aspect} 
 Aspect, A.   (2015) ``Closing the Door on Einstein and Bohr's Quantum Debate'', Physics 8, 123[https://physics.aps.org/articles/v8/123]. 
 
\bibitem{inference} Grangier, P.  (2021) ``Contextual inferences, nonlocality, and the incompleteness of quantum mechanics", 
 Entropy 23 (12), 1660; https://doi.org/10.3390/e23121660 
 
\bibitem{contextual} Grangier, P.   (2002) ``Contextual objectivity: a realistic interpretation of quantum mechanics'', European Journal of Physics 23:3, 331 [arXiv:quant-ph/0012122]. 

\bibitem{nayla} Farouki, N.  and Grangier, P. (2021) ``The Einstein-Bohr debate: finding a common ground of understanding?",  Found. Sci. 26, 97-101  [arXiv:1907.11267]  

 
\bibitem{myst}  Grangier,  P.  (2023) ``Revisiting quantum mysteries", in {\it The Quantum-Like Revolution: A Festschrift for Andrei Khrennikov}, A. Plotnitsky and E. Haven eds, Springer Cham [arxiv:2105.14448].

\bibitem{completing}  Grangier, P.  (2021) ``Completing the quantum formalism in a contextually objective framework'',  
Found. Phys.  51, 76 [arXiv:2003.03121]  

\bibitem{random} Auff\`eves, A. and P. Grangier, P. (2018) ``What is quantum in quantum randomness?'', 
Phil. Trans. R. Soc. A 376, 20170322  [arXiv:1804.04807]

\bibitem{Bohr} Bohr, N. (1935) ``Can Quantum-Mechanical Description of Physical Reality be Considered Complete?'', 
Phys. Rev. 48, 696. 
  
\bibitem{Gleason} Auff\`eves, A. and Grangier, P. (2020)  ``Deriving Born's rule from an Inference to the Best Explanation", Found. Phys. 50, 1781-1793 [arXiv:1910.13738].  
 
\bibitem{Uhlhorn} Auff\`eves, A. and Grangier, P. (2022) ``Revisiting Born's rule through Uhlhorn's and Gleason's theorems'',  Entropy 24(2), 199 https://doi.org/10.3390/e24020199

\bibitem{rings} Murray, F.J.  and von Neumann, J.   (1943) ``On Rings of Operators IV",  Ann. Math. 44, 716. 

\bibitem{infinite} von Neumann, J. (1939)``On infinite direct products",  Compos. Math. 6, 1–77,  \\
http://www.numdam.org/item?id=CM$\_1939\_\_6\_\_1\_0$

\bibitem{MP1} Van Den Bossche, M. and Grangier, P. (2023) ``Contextual unification of classical and quantum physics",  Found. Phys.  53:45  [arXiv:2209.01463]. 

\bibitem{MP2} Van Den Bossche, M. and Grangier, P. (2023)``Revisiting Quantum Contextuality in an Algebraic Framework’’, Proceedings of the DICE 2022 Conference, in J. Phys. Conference Series [arviv:2304.07757].

\bibitem{MP3} Van Den Bossche, M. and Grangier, P. (2023)``Postulating the Unicity of the Macroscopic Physical World'', Entropy 25(12), 1600. 

\bibitem{fry} 
Fry, E.S.  and Walther, T.  (2000) ``Fundamental tests of quantum mechanics", Adv. At. Mol. Opt. Phys 42, 1.
\end{thebibliography}
\end{document}